\documentclass{jetpl}
\usepackage{amsmath,amsfonts,epsfig,graphics,amssymb}

\usepackage{hyperref}
\hypersetup{colorlinks=true,citecolor=blue}

\lat
\title{Resolving infall caustics in dark matter halos}
\rtitle{Resolving infall caustics in dark matter halos}
\sodtitle{Resolving infall caustics in dark matter halos}

\author{
K.\, Dolag$^{a,b}$, 
A.\,D.\,Dolgov$^{c,d,e,f}$, and
I.\,I.\,Tkachev$^{c,g}$
}

\sodauthor{Dolag, Dolgov, Tkachev}

\address{$^a$University Observatory Munich, Scheinerstrasse 1, 81679 M¨unchen, Germany\\~\\
$^b$Max–Planck–Institut f¨ur Astrophysik, Karl–Schwarzschild–Strasse 1, 85748 Garching bei M¨unchen, Germany\\~\\
$^c$Laboratory of Cosmology and Elementary Particles, Novosibirsk State University, Pirogov street 2, 630090 Novosibirsk, Russia\\~\\
$^d$Dipartimento di Fisica, Universit`a degli Studi di Ferrara,  Polo Scientico e Tecnologico - Edicio C, Via Saragat 1, 44122 Ferrara, Italy\\~\\
$^e$Istituto Nazionale di Fisica Nucleare, Sezione di Ferrara,
Polo Scientiﬁco e Tecnologico - Ediﬁcio C, Via Saragat 1, 44122 Ferrara, Italy
\\~\\
$^f$Institute of Theoretical and Experimental Physics,
Bolshaya Cheremushkinskaya ul. 25, 117218 Moscow, Russia\\~\\
$^g$Institute  for Nuclear Research of the Russian Academy of Sciences,
Moscow 117312, Russia\\~\\
}

\abstract{
We have found that the phase-space of a dark matter particles assembling a galactic halo in cosmological N-body simulations has well defined fine grained structure. Recently accreted particles form distinctive velocity streams with high density contrast. For fixed observer position these streams lead to peaks in velocity distribution. Overall structure is close to that emerging in the secondary infall model.
}
\PACS{98.35.Gi,98.35.Mp,95.35.+d}

\newcommand{\be}{\begin{equation}}
\newcommand{\ee}{\end{equation}}

\begin{document}

\maketitle

\textbf{1. Introduction.} 

\begin{figure}[t]
\begin{center}
\includegraphics[width=0.9\textwidth]{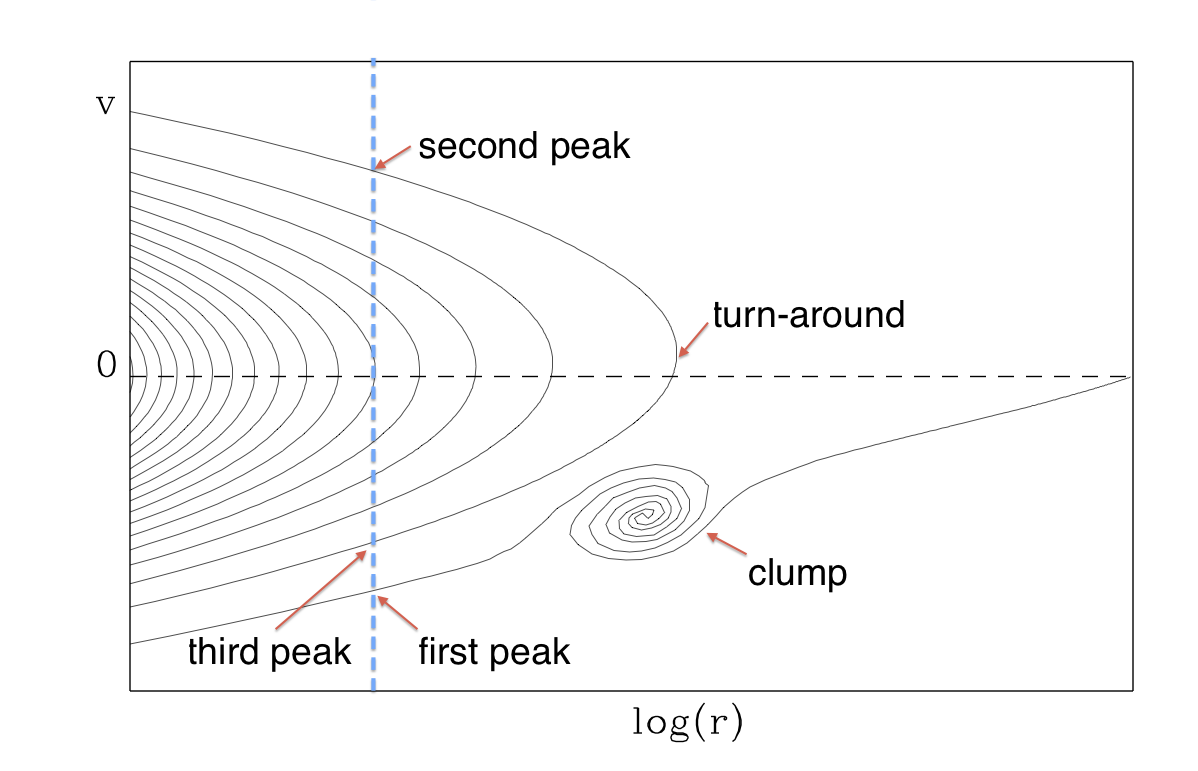}
\end{center}
\caption{Figure 1.  Phase space distribution of the halo dark matter particles at a fixed moment of time in the infall model. The solid curves represent occupied phase-space cells. The vertical dashed line corresponds to the observer position. Each intersection of the solid and dashed lines corresponds to a velocity peak in the velocity distribution measured by an observer. A small scale sub-clump falling into the galaxy for the first time is also  shown. Adapted from Ref.~\cite{Sikivie:1996nn}. \label{Fig:ssi}}
\end{figure}

Precise knowledge of phase-space distribution of the Galactic halo dark matter particles is important for the dark matter search experiments. 
Phase-space model, which is idealized but in certain aspects more realistic than e.g. traditional "isothermal" model and capable to describe fine-grained structure of the phase-space, was constructed in Refs.~\cite{Sikivie:1995dp,Sikivie:1996nn}. A distinctive feature of corresponding particle  distributions  is that the highest energy particles have discrete values of velocity~\cite{Sikivie:1992bk}. Qualitatively this conclusion is based on the following observation. Initial velocity dispersion of cold dark matter particles is very small, therefore one can consider that they occupy thin 3-dimensional
sheet in a 6-dimensional phase space, $\vec{v} = H\vec{r}$. Later on, when structure forms, the Hubble law would remain intact at large distances for isolated halos, but inside halos this sheet rolls and wraps around itself forming growing overdensity. Owing to the Liouville theorem the occupied phase-space sheet can neither cross itself nor be perforated. Resulting phase-space of the halo dark matter particles is depicted in Fig.~\ref{Fig:ssi}.

The model can be solved exactly with the assumption of self-similar, radial infall~\cite{Fillmore:1984wk,Bertschinger:1985pd}. In Ref.~\cite{Sikivie:1996nn} the model was extended to include angular momentum  in a self-similar way. (Phase-space shown in Fig.~\ref{Fig:ssi} corresponds to zero angular momentum.) With angular momentum included, the model explains not only flat rotational curves at intermediate distances, but also $\rho \propto r^{-\alpha}$ behavior of density, with $\alpha$ close to 1 in the cusps~\cite{Sikivie:1996nn}.  Note that assumptions of self-similarity and spherical symmetry are needed only to get exact solutions. With these assumptions relaxed, main qualitative features of the model would survive. In phase-space we would still have dense separated folds occupied by matter. Without self-similarity fold separations will change, while without spherical symmetry folds will became triaxial, but for a number of phenomenological applications this is inessential. For example, axion detectors have very good energy resolution and if dark matter is made of axions, such fine structure of velocity space can be resolved~\cite{Sikivie:1996nn}. 

Near the turn-around radius for each fold the density of dark matter in configuration space increases, such regions of space are called "caustics". It has been suggested  that caustics would increase the dark matter annihilation rate \cite{Bergstrom:2000bk,Mohayaee:2005fj, Mohayaee:2007pi} and that they can be probed by properly stacking the weak-lensing signal of a reasonable number of dark  halos~\cite{Mohayaee:2005jc}. 
An evidence for inner caustic rings distributed according to the predictions of the self-similar infall model has been found in the Milky Way \cite{Sikivie:1997ng,Sikivie:2001fg,Natarajan:2007xh}, in other isolated spiral galaxies \cite{Sikivie:1997ng,Kinney:1999rk}, and in galaxy clusters~\cite{Onemli:2007gm}, while there is clear evidence of continuing infall in our Local Group~\cite{Sandage,Steigman:1998sb} and also in the ensemble of halos in a wide range of scales from small galaxies to galaxy clusters~\cite{Boyarsky:2009af}.

Non zero primordial (e.g. thermal) velocities of dark matter particles can be accounted for analytically~\cite{Mohayaee:2005fj}. Their effect is negligibly small as far as dark matter is cold. In Ref.~\cite{Afshordi:2008mx} a formalism was developed which allows 
an analytical treatment of phase space streams and caustics without assumptions of
spherical symmetry, continuous or smooth accretion, and  self-similar infall for the formation of dark matter halos. 
However, the overall problem is difficult. Such infall picture will be destroyed 
e.g. by a major recent merger event. Influence of steady infall of small-scale clumps, as the one shown in Fig.~\ref{Fig:ssi} and which is outside of frameworks of self-similar infall model, is not clear also. Regarding influence of sub-clumps we would like to make several 
comments: i) Major fraction of dark matter resides outside of clumps and volume filling factor occupied by clumps is small. Outside of clumps the phase space may correspond to the infall model. ii) Clumps themselves follow infall trajectories in phase space. Inside host halo the clumps are tidally disrupted eventually. Released dark matter particles will continue to follow infall trajectories. The velocity dispersion in the resulting tidal streams may be smaller than in the clumps themselves since particles are released near the clump boundary were relative velocity is small. iii) The number of the observed dwarfs in our Galaxy halo is orders of magnitude less then the numerical simulations of $\Lambda$CDM model predict. A solution to this problem can be found on the assumption that the primordial power on small scales is reduced. Realization of this scenario in nature would bring realistic situation closer to 
an idealized infall model. 

Ultimately, one would like to resolve fine details of phase-space structure of collapsing halos in direct N-body approach.
Already in early simulations \cite{Zaroubi:1995dg} it was observed that the collapse of galactic halos in the cosmological setup proceeds gently and not via violent relaxation. A strong correlation of the final energy of individual particles with the initial one has been found indicating overall validity of infall picture. However, detection of caustic surfaces in N-body numerical modeling is a challenging problem. To understand and resolve the problem of fine structure of dark matter phase-space numerically, a dedicated very high resolution numerical simulations are needed. (To tackle the issue, novel numerical techniques are also emerging, superseding simple increase of the number of particles~\cite{Vogelsberger:2007ny,Vogelsberger:2009bn,Vogelsberger:2010gd}.)
Indeed, in simulations each "particle" exceeds by many orders of magnitude the solar mass, which is 60 to 80 orders of magnitude larger than the  mass of dark matter candidates being simulated.
This leads to artificial two body scattering, energy transfer and  to distorted orbits, see e.g. \cite{Diemand:2003uv}, while in reality the dark matter particles are
collisionless and pass unperturbed past each other.
Early simulations contained only a few thousand particles, and realistic study of phase space was hopeless. Kinematically cold infall streams were observed in more recent high resolution simulations. In Ref. \cite{Helmi:2002ss} it was found that the motions of the most energetic particles were strongly clumped and highly anisotropic. However, these simulations still produce rather smooth mass profiles without high density caustics and the mean contribution of an individual stream to local dark-matter distribution near the Sun position in the Galaxy was found to be of order $f_n \approx 0.3\%$ . This value of $f_n$ for the solar neighborhood is actually in agreement with predictions of self-similar infall model for $\epsilon = 1$, see Ref. \cite{Sikivie:1996nn}, where parameter $\epsilon$ defines the shape of initial overdensity, $\delta M(r_i)/M(r_i) = \left(M_0/M(r_i)\right)^\epsilon$.

Until recently, phase space picture qualitatively similar to the one presented in Fig.~\ref{Fig:ssi} was not reported from N-body simulations, see e.g. Ref.~\cite{Vogelsberger:2008qb}. However, in Ref.~\cite{Diemand:2008gf} it was shown that particle orbits in simulated cosmological cold dark matter halos are surprisingly regular and periodic, while the phase-space structure of the outer halo shares, qualitatively, some of the properties of the classical self-similar secondary infall model. Namely, the positions of six outermost resolved caustic  were in very good agreement with the infall model for $\epsilon = 1$. The ratios of caustic radii to their turnaround radii also agreed quite well with the model. However, reported trajectories in the radial velocity - radius plane 
were severely broadened producing only small ($<$ 10\%) enhancements in the spherical density profile.

In this paper we aim to reanalyze the phase-space structure of the outer halo. Our main point is based on the following observation. Realistic halos are triaxial and mass accretion is non-radial. In this situation the averaging over all directions in configuration space (as usually done) will not produce meaningful representation of 
the phase-space structure of two dimensional $(v_r,r)$ subspace. Mapping in this way, say, even infinitely thin ellipsoidal surfaces, will produce broad band distribution and real nature of phase-space sheets may be lost. To reveal it, one should try to display phase space with as small range for averaging, as statistics  allows. To verify this ~proposition we construct phase-space by doing 
averages in a sequence of decreasing solid angles around some arbitrarily chosen direction in configuration space. If phase-space becomes sharper in this sequence, then our proposition is correct.

\begin{figure}[t]
\begin{center}
\includegraphics[width=0.84\textwidth]{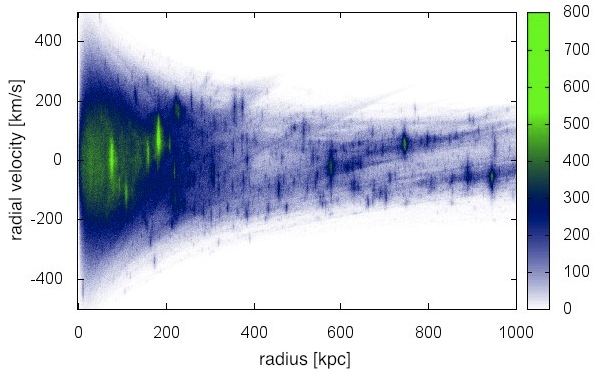}
\end{center}
\caption{Figure 2. Result of N-body simulation. Projection of phase-space of a  halo dark matter particles to $(r,v_r^{~})$ plane, being averaged over all angles in configuration space.  \label{Fig:phsp_vr}}
\end{figure}

\begin{figure}[t]
\begin{center}
\includegraphics[width=0.84\textwidth]{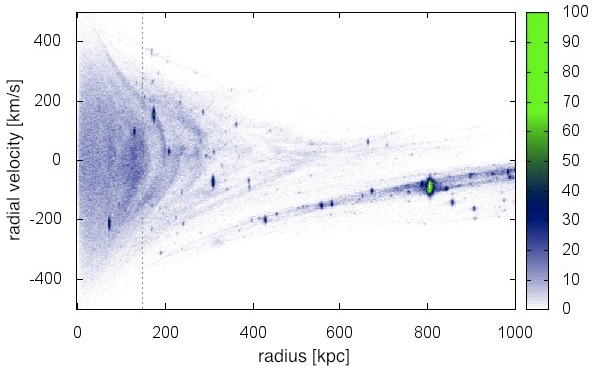}
\end{center}
\caption{Figure 3. The same as in Fig.~\ref{Fig:phsp_vr} but now averaging is restricted to solid angle  with opening  $\theta = 40^\circ$.  Dotted line represents ``an observer'' position at 150 kpc where the velocity distributions of Fig.~\ref{Fig:VD} are constructed.  
\label{Fig:phsp_vr20}}
\end{figure}

\textbf{2. Numerical sumulation.} 

We analyzed a N-body simulation of a galaxy size halo for which the
initial conditions were generated by Stoehr et al., Ref.~\cite{Stoehr:2002ht},
adopting a LCDM cosmology, with $\Omega_m = 0.3$, $\Omega_L = 0.7$,
$H_0 = 70$ km s$^{-1}$ Mpc$^{-1}$ and $\sigma_8 = 0.9$. This "Milky Way" like halo was
selected from an intermediate-resolution simulation as a halo with a maximum
rotational velocity at $z=0$ approximately equal to the Milky Way value
as a relatively isolated halo, which suffered its last major merger
at $z > 2$. We used the highest resolution initial conditions of that set, which
has a dark matter particle mass of $m_{dm} = 2\times 10^5~ M_\odot /h$, so that
the halo at $z=0$ was resolved by more than 11 million particles within the
viral radius. The total mass of the halo withing this radius therefore is
$M_{vir} = 2.3\times 10^{12}~ M_\odot/h$. The simulation was performd
using the cosmological N-Body code Gadget~\cite{Springel:2000yr,Springel:2005mi}
with the gravitational softening set to 0.2 kpc/h.

\textbf{3. Discussion.}

Phase-space of dark matter particles is 6-dimentional, while we have only 2 dimensions to display the results. We start with presenting it in $(v_r,r)$ subspace. In general non-spherically symmetric situation the occupation pattern of this subspace will depend upon direction chosen from the halo center. Having limited number of particles we cannot fix direction to a particular value, however. To accumulate statistics we simply integrate over all directions, as it is usually done. The result is shown in Fig.~\ref{Fig:phsp_vr}. We see host halo extending to about 300 kpc in radius and to 400 km/s in velocity.
We also see a  large number of infalling sub-clumps, but no strong evidence for infall trajectories. Even particles falling in for the first time are forming very wide band in the portion of phase-space between  400 kpc and 1 Mpc. Reason for this is clear: stacking even infinitely thin trajectories of the sort presented in Fig~\ref{Fig:ssi} will result in their dilution in overall ensemble if trajectories differ from each other.

To alleviate this problem we can sum up trajectories only in a limited range of viewing directions. To realize this procedure we choose arbitrary reference direction and sum occupation numbers of phase-space cells only if direction to a given cell lies within 
the cone with opening angle $\theta$ around the reference direction. 
The result is shown in Fig.~\ref{Fig:phsp_vr20} for $\theta = 40^\circ$.\footnote{"Hairy" tail in the region of distances form 300 kpc to 700 kpc may be a numerical  artifact. Similar distortions appear in idealized  spherically symmetrical situation without initial small-scale perturbations when the problem is solved using 1-dimentional N-body simulations instead of methods developed in Refs~\cite{Fillmore:1984wk,Bertschinger:1985pd}. If true, this indicates  that the processes of two body scattering are distorting phase-space in our simulation.}
Phase-space becomes much sharper now. First infall trajectory can be identified and several folds similar to those in Fig.~\ref{Fig:ssi} can be visually recognized in the host halo. This tendency continues with decreasing solid angle.

\begin{figure}[t]
\begin{center}
\includegraphics[width=0.87\textwidth]{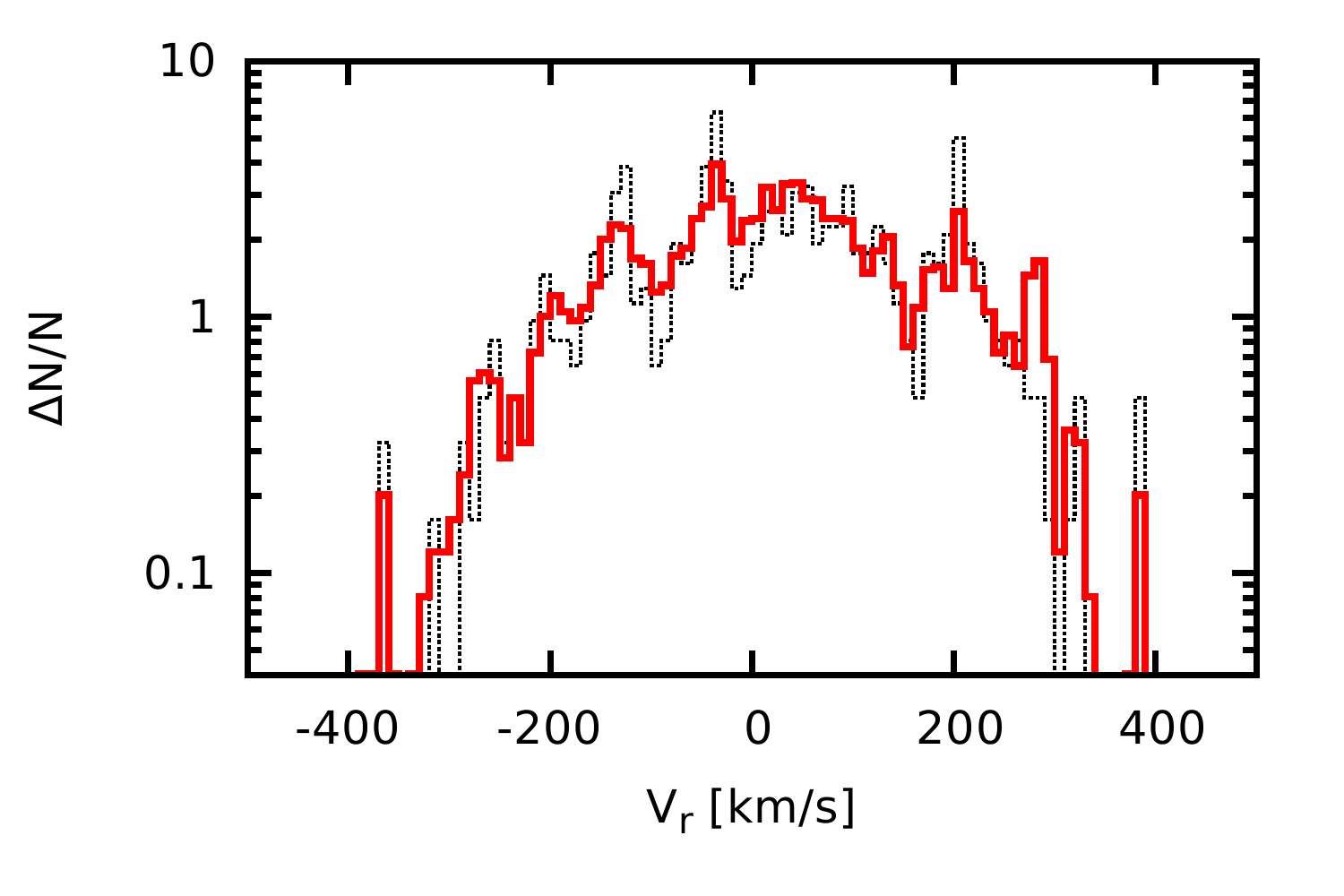}
\end{center}
\caption{Figure 4. Velocity distribution function at the observer position  $r = 150$ kpc.
Averages are done for fixed  $\Delta r = 2$~kpc and solid angles with openings  $\theta =40^\circ$  (thick solid line) and $\theta = 20^\circ$ (thin dotted line).  \label{Fig:VD}}
\end{figure}
\begin{figure}[t]
\begin{center}
\includegraphics[width=0.87\textwidth]{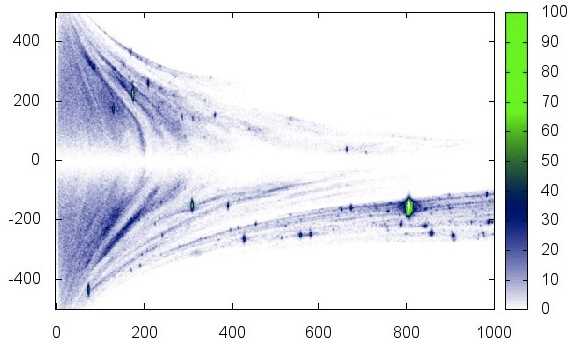}
\end{center}
\caption{Figure 5.  The same as in Fig.~\ref{Fig:phsp_vr20}. However, now 
$|\vec{v}|\cdot{\rm sign} (v_r)$ insted of $v_r^{~}$ is used. 
\label{Fig:phsp_vm}}
\end{figure}

To quantify the tendency we constructed velocity distributions at fixed observer position at $r = 150$ kpc for different values of $\theta$, these are presented in Fig.~\ref{Fig:VD}.  Thick solid line represents averaging inside solid angle with opening  $\theta =40^\circ$ while thin dotted line corresponds to $\theta = 20^\circ$. Peaks in the velocity distributions correspond to crossings of velocity streams in phase-space,  see Fig.~\ref{Fig:phsp_vr20}.  While amplitudes of peaks are subject to Poisson fluctuations  to some extent, and occasionally an amplitude may be influenced by the presence of a sub-clump in the volume of averaging, overall, these peaks and their average amplitude are real. This can be verified e.g. by comparing positions of peaks in Fig.~\ref{Fig:VD} with streams in Fig.~\ref{Fig:phsp_vr20} and by changing observer position, which we have  also done. 

Notably, first and second peaks, at $v \approx -400$~km/s and $v \approx 400$~km/s correspondingly, are well separated and isolated from the background already at $\theta =40^\circ$. We also see that with decreasing $\theta$ the average amplitude of peaks becomes consistently higher and they become more narrow. On the average, the
peak amplitudes rise by a factor of few.
This can be compared with small 10\% density contrast which was found in Ref.~\cite{Diemand:2008gf} and which we also observe at $\theta = 4\pi$ when averaging is done over all angles. This proves our proposition - "blurring" of phase-space  at large $r$  in the current N-body simulations is due to averaging, which is extraneous to many phenomenological applications. In reality the phase space is sharper  but we cannot evaluate at present to what extent. When we diminish the solid angle of integration even further, peaks continue to grow, but statistics decreases and noise starts to appear. We cannot reach the regime when the amplitude of the peaks becomes $\theta$-independent, this will require more extensive numerical simulation.

However, the phase-space becomes significantly sharper already in the current simulation if instead of $v_r^{~}$ the velocity modulus is chosen to display a two dimensional projection of the six dimensional phase space. In Fig.~\ref{Fig:phsp_vm} we present this by plotting $|\vec{v}|$ multiplied by the sign of $v_r^{~}$. Multiplication by sign of $v_r^{~}$ is needed to separate incoming and outcoming velocity streams. In this figure several largest velocity streams are separated from the "noise" completely. They stay separated and well defined up to a small distances of order of the  Sun position in the Galaxy. Relative mass fraction in these streams is small however.  The phase-space is blurred for the earlier folds. Again, to clarify the situation in these region one needs more extensive numerical simulation.

Why the phase-space is sharper in $|\vec{v}|$ as compared to $v_r^{~}$? The reason can be that $v_r^{~}$ is simply not a suitable coordinate in generic non-spherical situation. But whatever the reason, for a number of applications it is enough to show that a choice  of coordinates exists where phase streams are well defined and separated even after averaging. This is relevant,  for instance, for direct dark matter detection experiments, in particular for axion searches, where $|\vec{v}|$ is important, but not $v_r^{~}$.

We would like to comment also on the following peculiarity of Fig.~\ref{Fig:phsp_vm} related to  the empty band which is passing through the middle of the plot  along 
line $|\vec{v}| = 0$. It appears because $|\vec{v}|$ cannot be smaller than the value of 
the transverse component of the velocity near the turnaround point. Therefore, this portion of the phase-space is unoccupied, contrary to  $v_r^{~}$ projection. The boundary of this band immediately tells us  the value of transverse velocity and consequently the value of angular momentum $j$. The fact that boundary of the band stays parallel to $|\vec{v}| = 0$ axis tells that $j(r) \propto r$.
 
\newpage

\textbf{4. Conclusions.}  

We have demonstrated that the phase-space of a galactic halo emerging in cosmological N-body simulations has sharp fine grained structure. The density contrast is high at velocity streams, at least an order of magnitude  higher than previously reported. This is true for at least several latest streams, i.e. for those which correspond to recently accreted particles. Overall structure is close to the one emerging in the infall model. Our results can be  important for the dark matter search experiments.

{\bf Acknowledgements.}  We are indebted to S. I. Blinnikov, G. Rubtsov, O. Ruchayskiy,  P. Sikivie and P.~Tinyakov for useful discussions and comments. 

This work was supported by the Grant of the Government of Russian Federation (11.G34.31.0047). The work of I.T. has been supported in part by the SCOPES program.
K.D. acknowledges the support by the DFG Priority Programme 1177 and additional support by the DFG Cluster of Excellence "Origin and Structure of the Universe".

\end{document}